\documentclass[a4paper,12pt]{article}
\usepackage[warn]{mathtext}
\usepackage[T2A]{fontenc}
\usepackage[cp1251]{inputenc}
\usepackage[english]{babel}
\usepackage{amssymb,amsfonts,amsmath,mathtext,cite,enumerate,float}
\usepackage{graphicx} 
\usepackage{indentfirst}
\usepackage{psfrag}

\newcommand{\tocsecindent}{\hspace{6mm}}

\usepackage{geometry}
\makeatletter
\renewcommand{\@biblabel}[1]{#1.}
\makeatother

\begin{document}
\title{Melting Point Shift in Supported Metal Nanoclusters}
\author{V. D. Borman, P. V. Borisyuk, M. A. Pushkin, I. V. Tronin, V. N. Tronin,\\ V. I. Troyan, O. S. Vasiliev, M.V. Vitovskaya}
\date{}
\maketitle
\begin{abstract}
The dependency of the melting point of supported metal nanoclusters as function of clusters height is theoretically investigated in the framework of the uniform approach. The vacancy mechanism describing the melting point shift in nanoclusters with decrease of their size is proposed. It is shown that the essential role in clusters melting point shift is played by van der Waals forces of cluster-substrate interaction. It is shown, that the account of layer--by--layer fusion of a cluster allows to satisfactorily describe the melting of nanoclusters of  various metals, deposited onto a different substrates. The proposed model satisfactorily accounts for the experimental data.
\end{abstract}

One of the interesting properties of nanometer-sized systems is a shift of the melting point and the lattice parameter with decrease of cluster size (size effect). This effect was experimentally observed for Au, Ag and Cu\cite{Ph.Buffat1976,Gladkih1998en,Q.Xu2006,Tagaki1954,Bogomolov1976en,T.Castro1990} clusters deposited onto various substrates (W, C, Ni). The existing theoretical descriptions of the observed size effect are based on various approaches \cite{Ph.Buffat1976,Gladkih1998en,Pawlow1909,Skripov1981,M.Ya.Gamarnik1989, Morohov1979aen}. However, their common and essential drawback is that the influence of the substrate is not taken into account. At the same time it is experimentally proved \cite{T.Castro1990} that substrate plays a crucial role and the use of different substrates (W, C) results in the significant difference in melting points of clusters (up to 20\%  for 20\AA{} clusters on carbon and tungsten substrates  \cite{T.Castro1990}). Physically it can be related to the additional energy of the supported cluster due to the contact potential difference \cite{Lifshitz1984en}, the lattice mismatch at the cluster-substrate interface \cite{Devyatko2006en}, and the cluster-substrate van der Waals interaction \cite{Dzyaloshinskii1961en} that can lead to the shifts of cluster’s lattice parameter and melting temperature. Therefore, the understanding of the physical reason for the size shift of cluster’s melting point and lattice parameter needs the determination of the influence of the substrate on cluster’s properties.

In this paper the theoretical description of the melting point and the lattice parameter shifts for the supported nanoclusters of various metals carried out in the framework of the uniform approach using the vacancy mechanism taking into account the cluster-substrate interaction is proposed. In the framework of this mechanism the change of cluster properties with decrease of its size is described as a result of anharmonic oscillations of cluster atoms, leading to the decrease of the energy of vacancy formation and, hence, to the additional creation of vacancies in cluster with decrease of its size.

Recently in this paper\cite{Borman2009} the theoretical description of the melting point for the supported nanoclusters of various metals carried out in the framework of the uniform approach using the vacancy mechanism taking into account the cluster-substrate interaction is proposed.  In the framework of this mechanism the change of cluster properties with decrease of its size is described as a result of anharmonic oscillations of cluster atoms, leading to the decrease of the energy of vacancy formation and, hence, to the additional creation of vacancies in cluster with decrease of its size. 

In the paper \cite{Borman2009} the qualitatively agreement with experimental data of melting of Au clusters, deposited onto various substrates (W, C) is obtained. The difference may be related to the cluster-substrate lattice mismatch, to the contact potential difference at the interface, as well as to the fact that melting of the cluster doesn't occur in the whole volume of cluster but begins with the surface layers while the inner layers remain solid (that means the existing both the liquid surface layer and solid core, ``layer-by-layer melting'' below). The letter effect is observed experimentally \cite{T.Castro1990, david1995, kofman1994}. Thus, the quantitative description of the melting temperature shift needs the determination of the influence of these effects on cluster's properties.

In this paper the influence of the contact potential difference at the interface, the finite lateral  cluster size and the layer-by-layer melting of the clusters had been taken into account so that we succeeded in quantitative description of the melting temperature shift of different metal clusters deposited onto different substrates (Au, Ag clusters deposited onto surfaces of W, HOPG, Ni).

This effect can be described considering the van der Waals interaction of a supported metal cluster with bulk metal or semimetal substrate \cite{Dzyaloshinskii1961en}. Van der Waals forces are of the electromagnetic character and arise due to the mutual polarization of neutral atoms of the interacting objects. The presence of these forces can lead to the change of the chemical potential of a cluster. In this case, if the change is positive, it can be interpreted as an additional thermodynamic potential per atom of a cluster, arising due to the interaction with substrate. It results in the decrease of the binding energy of atoms in a lattice and, therefore, in the increase of the probability of vacancies formation in a cluster. Besides van der Waals forces, the formation of vacancies in a cluster occurs due to the presence of cluster surface, that results in the dependence of the energy of vacancy formation on its distance from cluster surface, as well as on cluster shape \cite{Devyatko1990ben}. Under assumption that the clusters under investigation \cite{Ph.Buffat1976, Gladkih1998en, Q.Xu2006, Tagaki1954, Bogomolov1976en} satisfy the requirement   (where  $l=0.7-10.0$ nm and  $h=0.3-5.0$ nm are cluster’s lateral size and height, consequently), the cluster surface can be considered as a flat and the influence of the boundary effects related to cluster’s shape can be neglected.

The equilibrium concentration of vacancies in a cluster can be found from the requirement of the minimum of the free energy of cluster’s vacancies subsystem taking into account the interaction with substrate:

\begin{equation}
\label{eq:math:free_en}
\left. {\frac{\delta F}{\delta n}} \right|_{n=\overline{n}}  = 0
\end{equation}

where $\overline{n}$ is the average concentration of vacancies in a cluster. The interaction of vacancies at the distance more than the character size of the vacancy (the possibility of divacancy formation) is attractive, while at the distance less than the characteristic size of the vacancy it has a character of a solid core (it is impossible to locate two vacancies at the same site of the lattice). Assuming the equilibrium concentration of vacancies much less than the atomic density in a cluster, the vacancies subsystem of a cluster can be considered as a van der Waals’ gas \cite{Devyatko1990aen}. In this case the free energy of the vacancies subsystem after decomposition for virial coefficients is given by \cite{Landau1980en}:

\begin{equation}
\label{eq:math:svob_en1}
F = F_{id}  + \frac{N^2 TB\left( T \right)}{V} \,;\quad B(T) = \frac{1}{2}\int {\left( {1 - \exp \left( { - \frac{U_{12}}{T}} \right)} \right)} \,dV
\end{equation}

where $N$ is the number of vacancies in a cluster, $V$ is the cluster volume, $U_{12}$  is the interaction potential of vacancies, and $F_{id}$ is the free energy of non-interacting vacancies system. Considering  $U_{12}$ as a rectangular potential well of depth  $E_{d}$ (the energy of divacancy formation), the virial coefficient  $B(T)$ is expressed as:

\begin{equation}
\label{eq:math:vir_koef}
B\left( T \right) = b + 7b\left( {1 - \exp \left( {E_d/T} \right)} \right)
\end{equation}

\noindent where $b = \frac{2\pi r_0^3}{3}$, $r_0$  --- is the characteristic size of a vacancy. The free energy $F_{id}$   of non-interacting vacancies system is given by \cite{Devyatko1990ben}:

\begin{equation}
\label{eq:math:svob_en_nevz_vac}
\begin{gathered}
F_{id} = \int {E_v} \overline{n} \,dV + T\int {\left[\overline{n}\ln {{\overline{n}/{n_0}}-\overline{n}}\right] \,dV}  \\
E_v = E_v^B  - \delta \mu^{vdv}+\delta \mu_c 
\end{gathered}
\end{equation}

where $E_B$, $E_S$ are the energies of vacancy formation in the volume and at the surface of a cluster, consequently, $n_0$  is the atomic density for a cluster, $a$  is the cluster lattice constant, $\delta \mu^{vdv}$  is the additive chemical potential of a cluster arising due to its interaction with substrate \cite{Dzyaloshinskii1961en}, $\delta\mu_c$ --- the shift of the cluster chemical potential due to the presence of the cluster-substrate interface. 

The energy of the vacancy formation $E_v^B$ is assumed depending on its distance from cluster surface\cite{Devyatko1990ben}: 

\begin{equation}
\label{eq:energ}
E_v^B  = E_B  - \left( {E_B  - E_S } \right)\exp \left( { - \frac{x}{2a}} \right)
\end{equation}

\noindent where $E_B$, $E_S$ are the energies of vacancy formation in the volume and at the surface of a cluster, consequently, $h$ is the cluster height, $a$  is the cluster lattice constant, $x$ is the distance from the cluster surface. Substituting the expression \eqref{eq:math:svob_en_nevz_vac}  and into \eqref{eq:math:svob_en1}, the free energy is obtained:

\begin{equation}
\label{eq:math:F}
F = \int {E_v \overline{n} \,dV}  + T\int \left[ \overline{n} \ln {{\overline{n}/{n_0 }} - \overline{n}} \right] \,dV  - nT \ln \left({1 - nb} \right) - 7n^2 bT\left({1 - \exp \left({E_d/T} \right)} \right)
\end{equation}

The equation for the equilibrium concentration of vacancies $\overline{n}$  in a cluster follows from \eqref{eq:math:F} and \eqref{eq:math:free_en} as:

\begin{equation}
\label{eq:math:ur_vac}
E_v^B  - \delta \mu^{vdv} + \delta \mu_c + T\ln \frac{\overline{n}}{n_0 } - T\ln \left( {1 - \overline{n} b} \right) + \overline{n} T\frac{b}{1 - \overline{n} b} - 14\overline{n} bT \left( {1 - \exp \left( {\frac{E_d }{T}} \right)} \right) = 0
\end{equation}

where  $\delta \mu^{vdv}$  can be calculated if the complex dielectric constants of substrate and cluster are known \cite{Dzyaloshinskii1961en}. The effect of the cluster surface taken into account in the expression for the energy of vacancy formation \eqref{eq:math:svob_en_nevz_vac}, the value $\delta \mu^{vdv}$ can be calculated for a film of thickness $h$ deposited onto substrate \cite{Dzyaloshinskii1961en}:

\begin{equation}
\label{eq:math:delta_mu}
\delta \mu  = \frac{a^3}{z}\frac{\hbar \overline{\omega}}{8\pi ^2 h^3 }\,;\quad \overline{\omega} = \int\limits_0^\infty  {\frac{{\left( {\varepsilon_f \left( {i\xi } \right) - 1} \right)\left( {\varepsilon_f \left( {i\xi } \right) - \varepsilon_s \left( {i\xi } \right)} \right)}}{{\left( {\varepsilon _f \left( {i\xi } \right) + 1} \right)\left( {\varepsilon_f \left({i\xi }\right) + \varepsilon_s \left( {i\xi } \right)} \right)}}} \,d\xi 
\end{equation}

where  $i\xi=\omega$  is the imaginary part of the electromagnetic field, $\varepsilon_s$, $\varepsilon_f$  are the dielectric constants of substrate and cluster materials, consequently, and  $z$ is the number of the nearest neighbors in a cluster. Under assumption of the validity of the expression $\varepsilon_j \left({i\xi }\right) = 1 + 4\pi k \frac{{\sigma_j }}{\xi}$, $j = s,f$  for the dielectric constants  \cite{Korneev, Devyatko1990ben}, $k=8.98\cdot 10^9 N\cdot m^2/C^2$ the value $\delta \mu^{vdv}$  is given by:

\begin{equation}
\label{eq:math:delta_mu_itog}
\delta \mu^{vdv}  = \frac{a^3k }{z}\frac{{\hbar (\sigma_f  - \sigma_s )}}{4\pi h^3 } \cdot \frac{\sigma_f }{\sigma_s } \cdot \ln \left( {\frac{\sigma_f  + \sigma_s }{\sigma_f }} \right)
\end{equation}

where $\sigma_f$, $\sigma_s$ are the conductivity of cluster and substrate, consequently.

The shift of the cluster chemical potential due to the presence of the cluster-substrate interface is given by \cite{Lifshitz1984en}: 

\begin{equation}
\label{eq:mu_c}
\begin{gathered}
\delta \mu _c  = A\exp \left( { - \kappa (h - x)} \right) \\
A = {\frac{\kappa _2 \left( {\mu _2  - \mu _1 } \right)}{\left( {\kappa _1  + \kappa _2 } \right)}}\\
\kappa _{1,2}^2  = 4\pi e^2 {\frac{n_{1,2}^{( 0)} }{kT}}
\end{gathered}
\end{equation}

where  $\mu_1$, $\mu_2$ are the chemical potentials of graphite (1) and Au (2), $n_{1,2}^{(0)}$  is the electron density.

\begin{figure}[h]
\psfrag{n}{$\displaystyle \frac{\overline{n}}{n_{0}}$}
	\centering
		\includegraphics[width=0.60\textwidth]{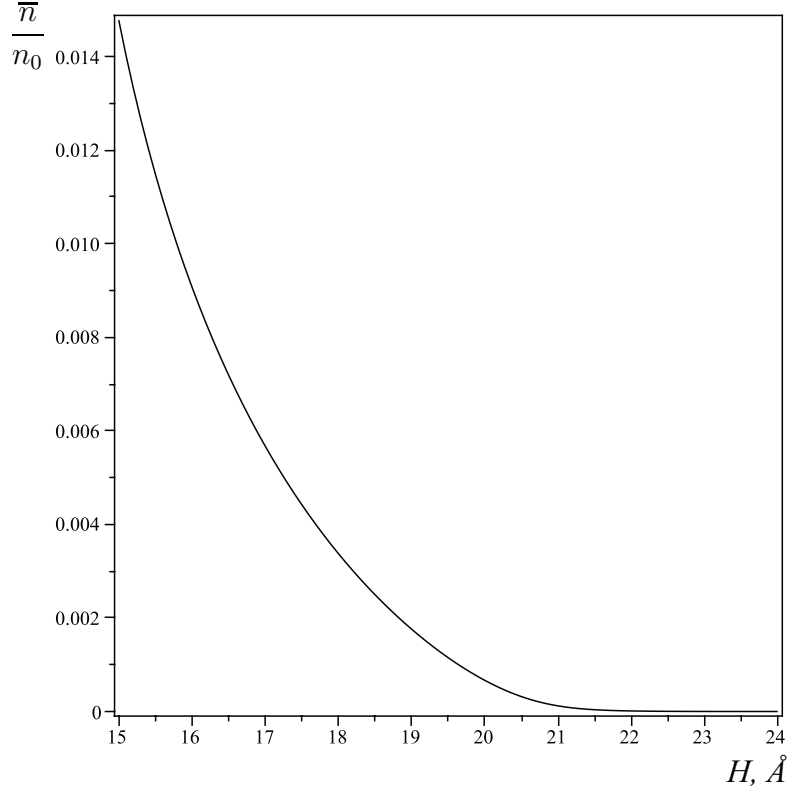}
	\caption{\small{The equilibrium concentration of vacancies  $\overline{n}$ in Au cluster on graphite at $T=300^{\circ} K$ relative to the atomic density  $n_{0}$ calculated as function of cluster height $h$. }}
	\label{fig:vac}
\end{figure}

The conductivity $\sigma_f$  of a cluster can be calculated by taking into account the contributions to the intrinsic resistivity of the cluster material $\rho_0$  and the additional resistivity  $\rho_{vac}$ arising due to the additional scattering of conduction electrons from vacancies. The additional resistivity $\rho_{vac}$ depends on the concentration of vacancies defined by the additional chemical potential arising due to the presence of a substrate. Thus, the expression for the cluster conductivity takes the form:

\begin{equation}
\label{eq:math:sigma}
\sigma_f  = \left({\rho_0+\rho_{vac}} \right)^{-1} \,;\quad \rho_{vac} = \rho_{vac}^{(0)} \cdot {{\overline{n}}/{n_0}}
\end{equation}

where  $\rho^{(0)}_{vac}$  is the additional resistivity per a vacancy that does not depend on  $\overline{n}$. The expressions \eqref{eq:math:ur_vac}, \eqref{eq:math:delta_mu_itog}, \eqref{eq:math:sigma} allow to calculate the equilibrium concentration of vacancies in a cluster   taking into account the cluster-substrate interaction. The theoretical dependence of the relative concentration of vacancies for Au cluster on graphite surface as function of cluster height at room temperature (Т=300$^{\circ} K$) calculated from the equations \eqref{eq:math:ur_vac}, \eqref{eq:math:delta_mu_itog} and \eqref{eq:math:sigma} is plotted in Fig.~\ref{fig:vac} . 

It is seen that the concentration of vacancies sharply increases for cluster height less than 22\AA, that can result in the modification of the modulus of elasticity of cluster and, as a consequence, to the shifts of the lattice parameter and the melting point \cite{Devyatko1990aen}.

The melting point can be calculated taking into account the dependence of the modulus of elasticity on the concentration of vacancies in a cluster \cite{Devyatko1990aen}. Using the vacancy model of melting \cite{Devyatko1990aen} and considering vacancies as dilatation centers, it is possible to calculate the shear modulus   as a function of the concentration of vacancies \cite{Devyatko1990aen}:

\begin{equation}
\label{eq:math:mod_K}
\widetilde{K}= K - \frac{4}{15}\pi^4 \left({K\Delta V} \right)^2 \frac{\overline{n}(1 - \overline{n}/{n_0 })^2 }{T\left[1+7{\bar{n}/{n_0 }}\left( 1 - {\overline{n}/{n_0}}\right)^2 \left({1-\exp \left({{{E_d}/T}}\right)}\right) \right]}
\end{equation}

Here $K$  is the shear modulus of the defect free material and $\Delta V$ is the dilatation volume.

Taking into account the finite lateral cluster size, the dilatation volume depending on the particle size can be found. The stress tensor $ \sigma _ {ik} ^ {td} $ in an elastic medium with point defects can be calculated in the model considering the defect as centre of a dilatation \cite {Eshelby1956en}:

\begin{equation}
\label{eq:sigma_td}
\sigma _{ik}^{td}  = \left\{ {\left( {K - {\frac{2}3}\mu } \right)\delta _{ik} \delta (\vec{ r }) + 2\mu {\frac{\delta _{ik} }{r^3 }} - 6\mu {\frac{x_i x_k }{r^5 }}} \right\}\Delta V
\end{equation}

\noindent where $\mu$ is the modulus of dilatation.

On the other hand, let's consider a spherical cluster with point defect as a hollow ball with exterior radius $R_1$, interior radius $R_2$ with pressure outside $p_1$, and pressure in a cavity $p_2=0$. In this case strain of a hollow ball is given by \cite {Landau1986en}:

\begin{equation}
\label{eq:ui}
u_i  = \left( {a + {\frac{b}{r^3 }}} \right)x_i \,,
\end{equation}

\noindent where $ a = {\frac{1}{3K}}\left[ {{\frac{p_2 R_2^3 }{R_1^3  - R_2^3 }}} \right] $  and $b = {\frac{p_2 }{4\mu }}\frac{R_2^3 R_1^3 }{R_1^3  - R_2^3 }$ has been found from  boundary conditions \cite{Landau1986en}. In this case the stress tensor  $\sigma _{ik}$ takes the form: 

\begin{equation}
\label{eq:sigma}
\sigma _{ik}  = \left\{ {\left( {{\frac{1}{3}}\left( {K - {\frac{2}{3}}\mu } \right)\delta \left( {\vec{r} } \right) + 3Ka} \right)\delta _{ik}  + 2\mu {\frac{\delta _{ik} }{r^3 }} - 6\mu {\frac{x_i x_k }{r^5 }}} \right\}b
\end{equation}

\noindent Equations \eqref{eq:sigma_td} and \eqref{eq:sigma} allows to calculate the dilatational volume $\Delta V$ as a function of the cluster radius:

\begin{equation}
\label{eq:dv}
\Delta V = b = {\frac{p_2 }{4\mu }}{\frac{R_2^3 R_1^3 }{R_1^3  - R_2^3 }}
\end{equation}

The dilatational volume $\Delta V_0$ for a continuous medium with a radius of cavity $R_2$ can be found from boundary conditions ($ R_1 \to \infty $, $ \Delta V \to \Delta V_0$):

\begin{equation}
\label{eq:dv0}
\Delta V_0  = {\frac{p_2 }{4\mu }}R_2^3 \, ,
\end{equation}
 
\noindent Finally the dilatation volume $\Delta V$ takes the form:

\begin{equation}
\label{eq:dv_it}
\Delta V = \Delta V_0 {\frac{R_1^3 }{R_1^3  - R_2^3 }}
\end{equation}

\noindent Substituting the expression \eqref{eq:dv_it} into \eqref{eq:math:mod_K} the shear modulus is obtained:

\begin{equation}
\label{eq:K_it}
\widetilde{ K} = K - \frac{4}{15}\pi ^4 \left( {K\Delta V_0 \frac{R_1^3}{R_1^3  - R_2^3 }} \right)^2 {\frac{\overline n \left( {1 - {{\overline n } \mathord{\left/
 {\vphantom {{\overline n } {n_0 }}} \right.
 \kern-\nulldelimiterspace} {n_0 }}} \right)^2 }{T\left[ {1 + 7{{\bar n} \mathord{\left/
 {\vphantom {{\bar n} {n_0 }}} \right.
 \kern-\nulldelimiterspace} {n_0 }}\left( {1 - {{\overline n } \mathord{\left/
 {\vphantom {{\overline n } {n_0 }}} \right.
 \kern-\nulldelimiterspace} {n_0 }}} \right)^2 \left( {1 - \exp \left( {{{E_d } \mathord{\left/
 {\vphantom {{E_d } T}} \right.
 \kern-\nulldelimiterspace} T}} \right)} \right)} \right]}}
\end{equation}

\begin{figure}[H]
\psfrag{K}{$\displaystyle \frac{\widetilde{K}}{K}$}
\centering
\includegraphics[width=0.5\linewidth]{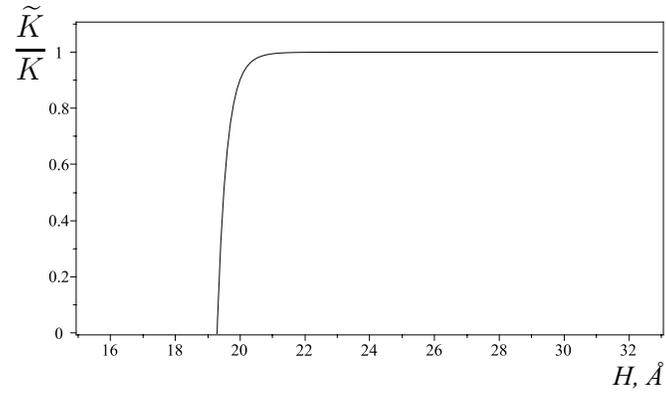}
\caption{\small{The dependence of a shear modulus $ \widetilde {K}/K $, where $K $ is a shear modulus of defect-free substance on the cluster height $H $ at temperature $T=300 ^ {\circ} K $}} 
\label{fig:mod_k}
\end{figure}

\begin{figure}[H]
\begin{minipage}[h]{0.47\linewidth}
\center{\includegraphics[width=1\linewidth]{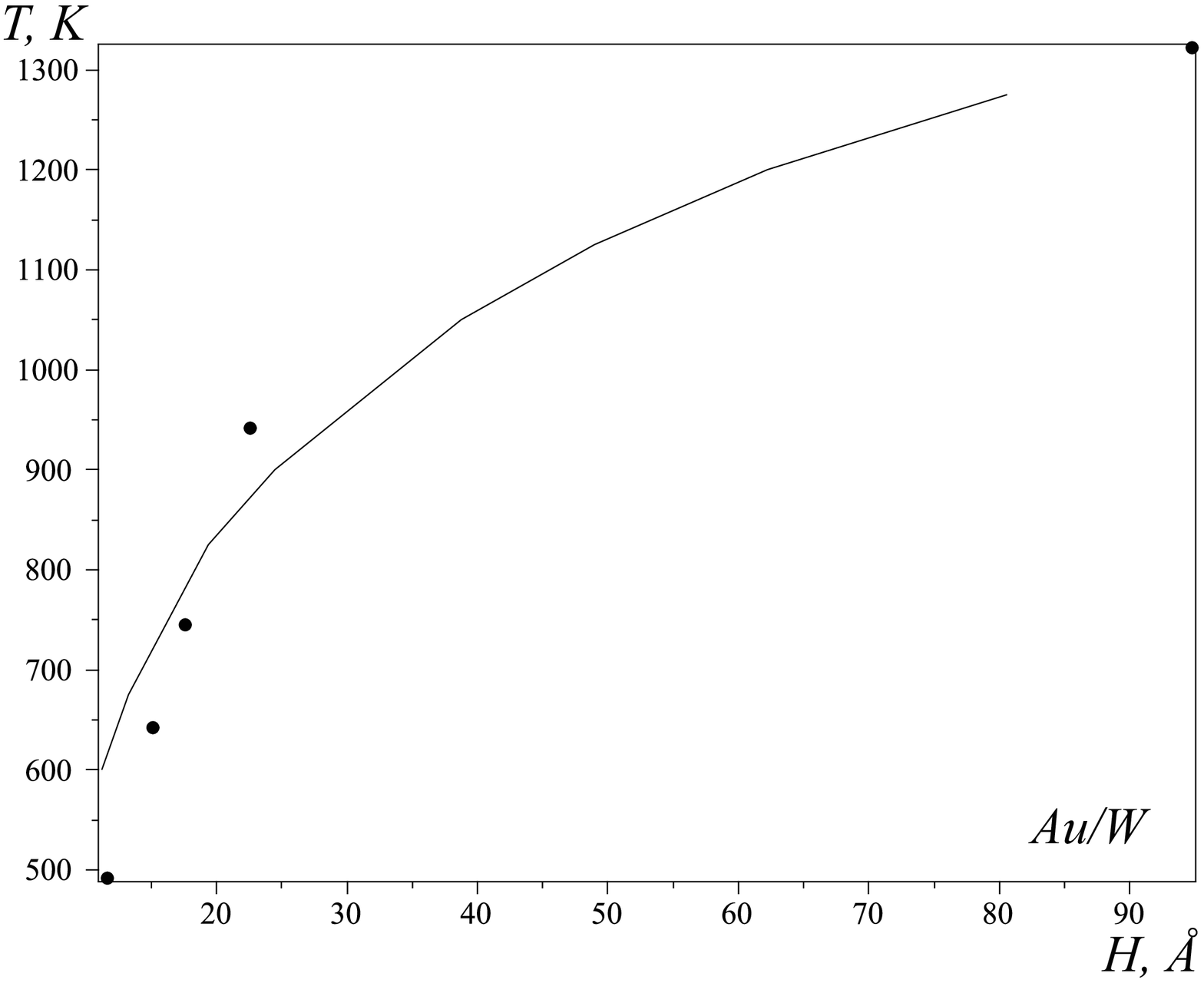}}
\end{minipage}
\hfill
\begin{minipage}[h]{0.47\linewidth}
\center{\includegraphics[width=1\linewidth]{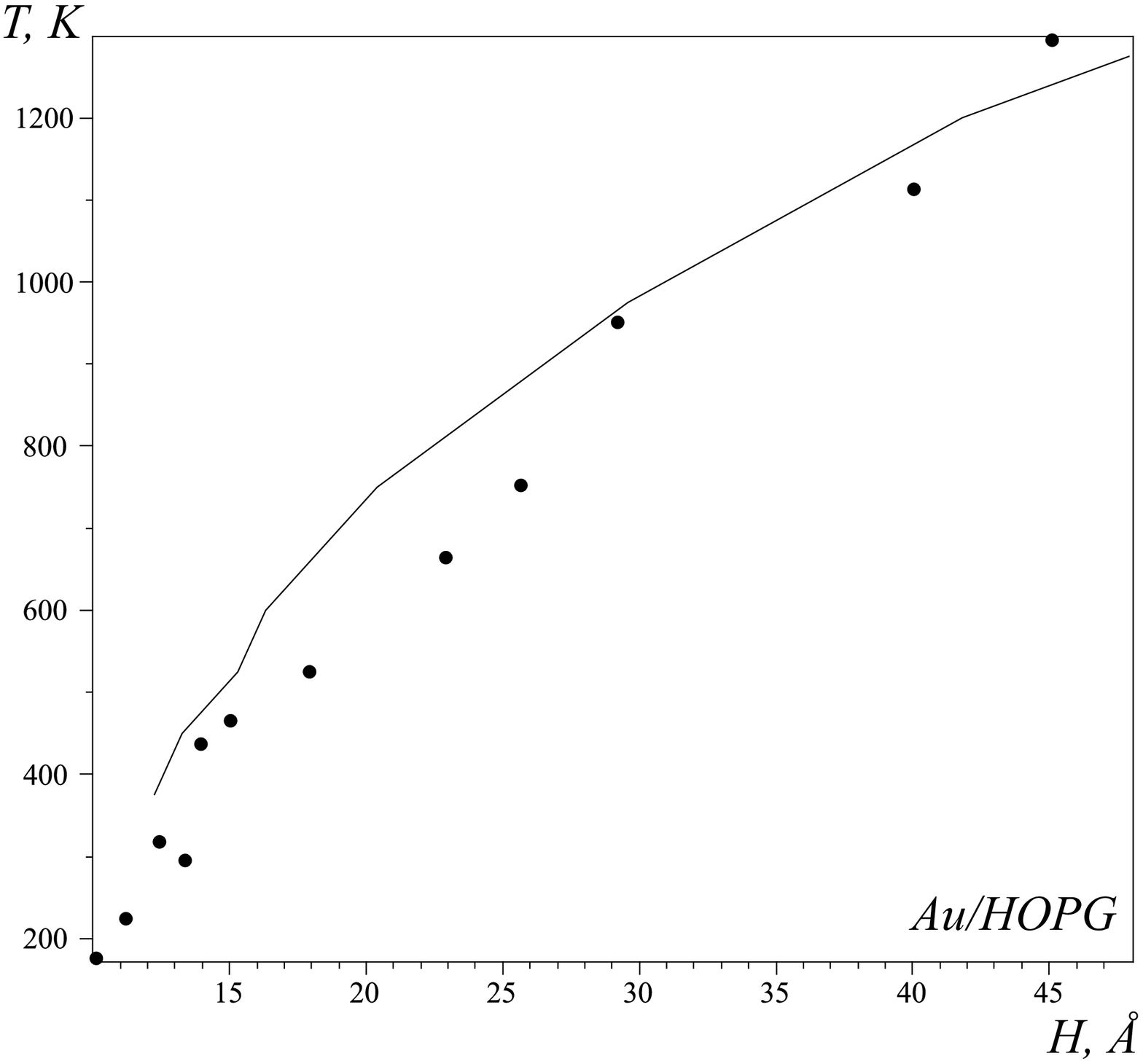}} 
\end{minipage}
\vfill
\begin{minipage}[h]{0.47\linewidth}
\center{\includegraphics[width=1\linewidth]{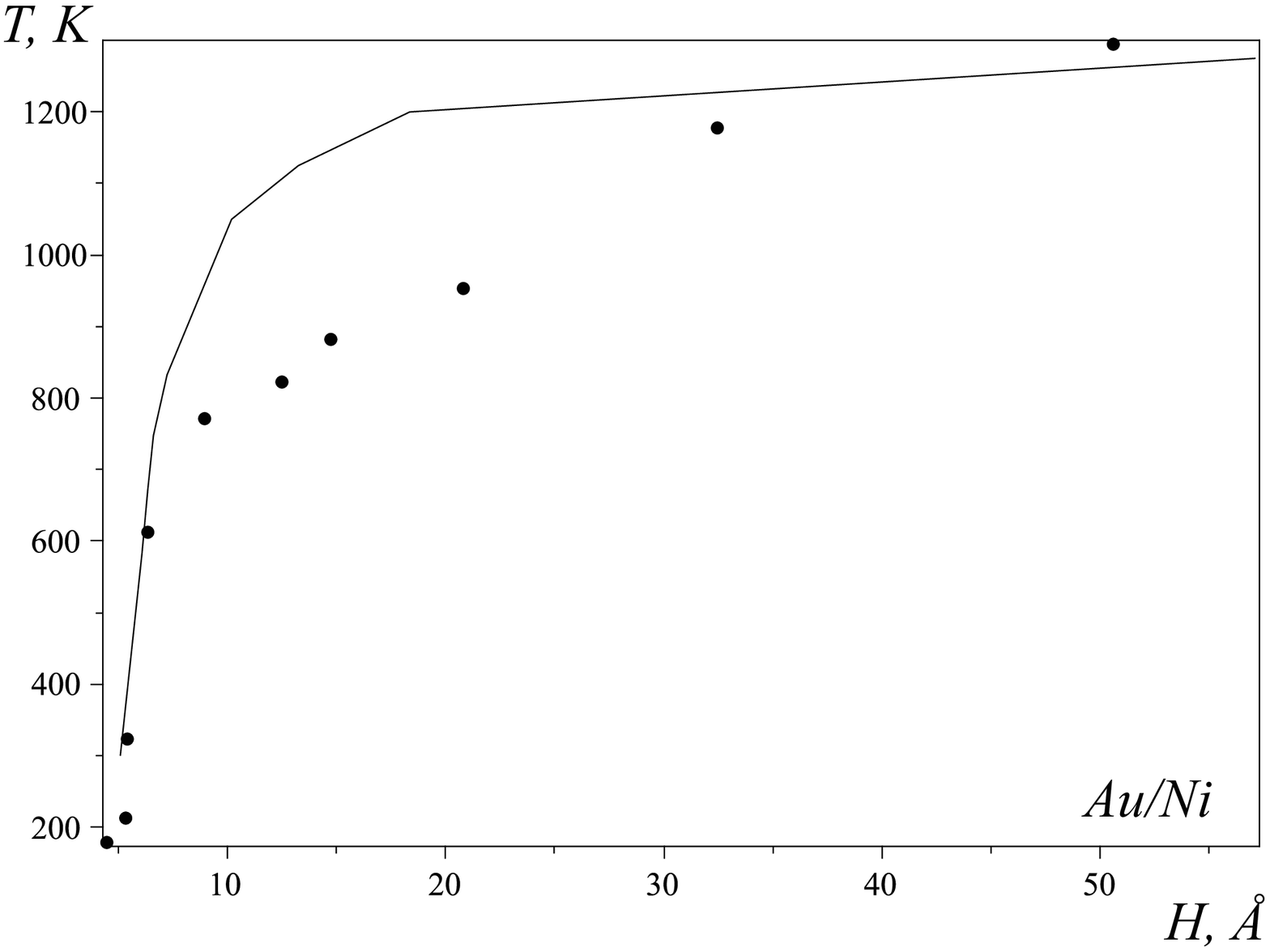}} 
\end{minipage}
\hfill
\begin{minipage}[h]{0.47\linewidth}
\center{\includegraphics[width=1\linewidth]{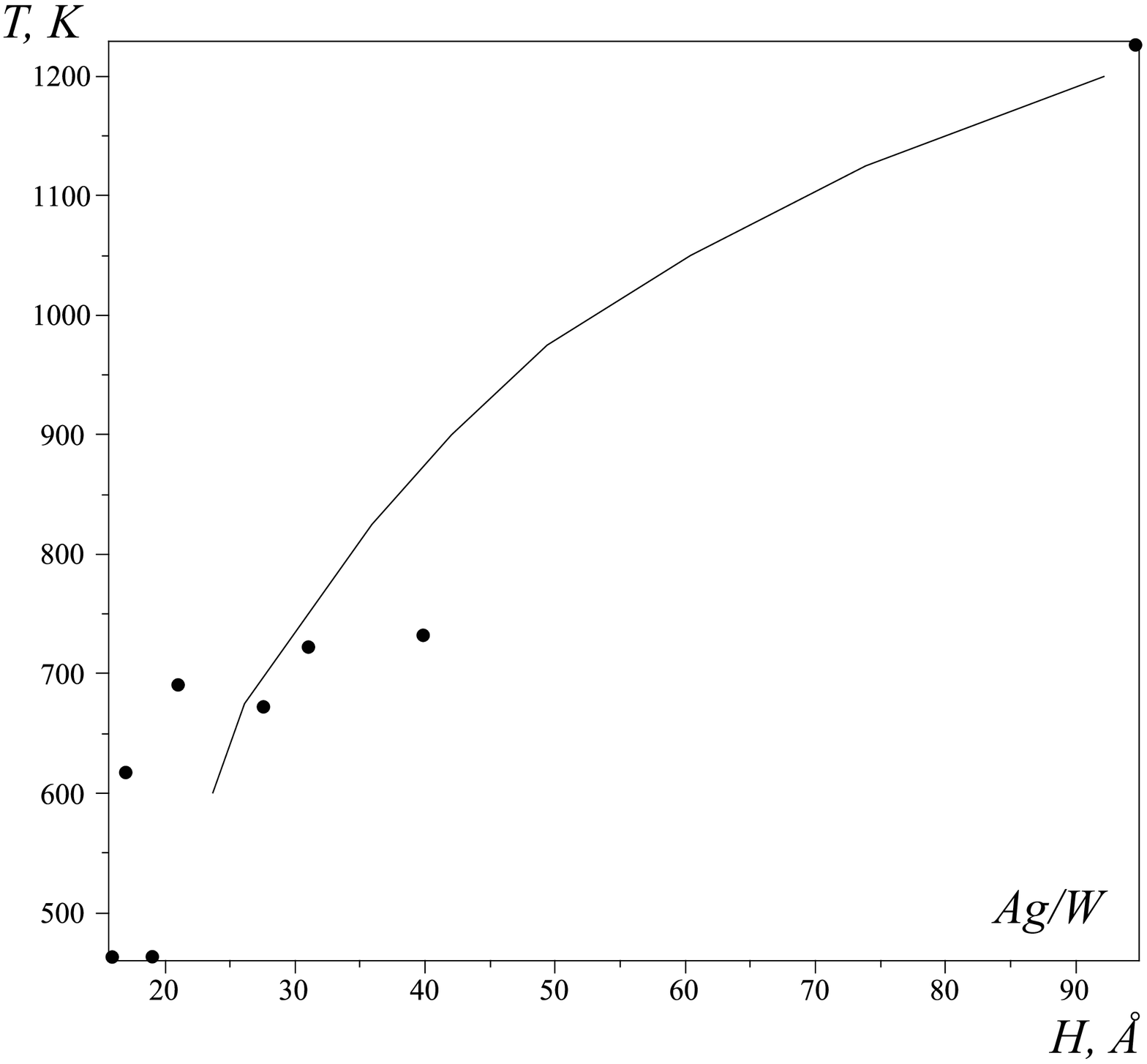}}
\end{minipage}
\caption{\small{The dependence of the melting point  $T_{melt}$ of different clusters on its height  $H$. Solid curves are the theoretical dependences calculated using equations \eqref{eq:K_it} for Au, Ag clusters on W, HOPG, Ni. Points are the experimental data (Au/HOPG, Au/Ni \cite{Borman2010}, Au/W, Ag/W \cite{T.Castro1990}) }}
\label{fig:tempplav}
\end{figure}

The dependence of the shear modulus $ \tilde {K} $ of Au clusters supported on carbon sas a function of cluster height $L$ at temperature $T=300 ^ {\circ} $ is presented in Fig.~\ref {fig:mod_k}. Defining the melting point of a cluster from the Born’s criterion $\tilde{K}(T_{melt})=0$ \cite {Devyatko1990aen}, the dependence of the melting temperature of a layer at a distance $x $ from a cluster surface on cluster height can be obtained from the equation \eqref {eq:math:mod_K}. The cluster is considered to be melted if it has no solid core. The dependencies of the melting temperature of the last layer (the nearest to the cluster - a substrate interface) for Au and Ag clusters deposited onto various substrates (HOPG, W, Ni)  as function of cluster height, calculated from \eqref {eq:K_it} are presented in Fig.~\ref {fig:tempplav}.

It is seen from Fig.~\ref {fig:tempplav} that an additional formation of vacancies in a cluster due to the interaction with substrate, leads to reduction of melting temperature that quantitatively agrees with the experimental data. 

\begin{figure}[H]
\centering
\includegraphics[width=0.6\linewidth]{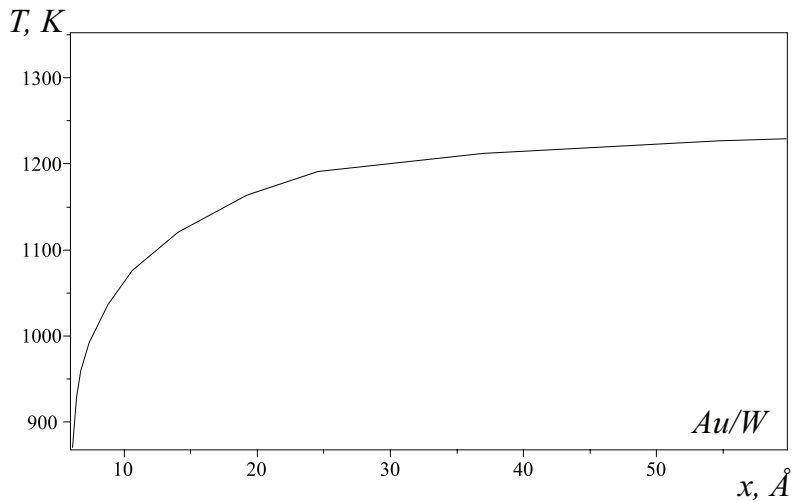}
\caption{\small{The dependence of melting temperature of the layer which as a function of distance $x $ to the surface of Au cluster of $H=60$\AA{} in  height, deposited onto W surface}} 
\label{fig:ag_w_layer}
\end{figure}

Let's consider the dependence of melting temperature of a cluster on distance to a cluster surface. The theoretical dependence of melting temperature of layers of Au cluster of $H=60$\AA in height, deposited onto W surface, as a function of a distance to the cluster surface is presented in fig.~\ref {fig:ag_w_layer}.

It is seen from fig.~\ref {fig:ag_w_layer}, that melting temperature of surface layers is considerably lower ($ \approx$1.5 times), than melting temperature of the layers neighboring   to a cluster -- substrate interface. Surface of a cluster  of 60\AA {} in height is considered to be melted already at $T=800 ^ {\circ} K $ whereas the inner layers are melted at considerably higher temperatures, so that the melting of a whole cluster with height of 60\AA {} occurs at temperature $T\approx 1200 ^ {\circ} K $. This behavior is observed experimentally for various materials of clusters (see for example \cite {T.Castro1990, david1995, kofman1994}). \\

Let's estimate contributions of the considered effects to the melting temperature of clusters: the additional energy of a cluster per one atom due to its interaction with substrate can be estimated in the frames of several mechanisms taking into account the contact potential difference, the lattice mismatch at the interface, the van der Waals interaction with the substrate, as well as the layer-by-layer melting.

The estimations will be carried out for the cluster of 30\AA{} in height. According to the equations \eqref{eq:math:delta_mu_itog} and \eqref{eq:math:sigma} , the additional energy of a cluster per one atom due to the van der Waals interaction with substrate is $\left. {\delta \mu _{cluster}^{VdV} } \right|_{h = 30\text{\AA}} \approx 0.2eV$.

The additional energy of a cluster obtained due to the contact potential difference can be estimated assuming that the influence of a substrate is considerable only within the first monolayer of atoms at the interface: $\delta \mu_c \approx \frac{1}{V}\int {\delta \mu} \,dV \approx \frac{{\delta \mu \pi r^2 {1/ \kappa }}}{\pi r^2 h}\approx \frac{{\delta \mu }}{{h\kappa }}$ , where  $V$, $h$ are the volume and the height of a cluster. In this case the value of the additional energy due to the contact potential difference equals to $\left. {\delta \mu_{cluster}^\phi}\right|_{h = 30\text{\AA}} \approx -0.05eV$.

The additional energy of a cluster per one atom due to the lattice mismatch at the interface can be estimated considering that the characteristic temperature of the transition to the disproportionate phase is about $\approx 500^{\circ}K$ \cite{Galitsyn1998en}. Thus, the additional energy per one atom due to the this mechanism is  $\delta \mu_{cluster}^{lattice} \approx 0.05eV$.

In this paper the energy of vacancy formation had been taken into account to be different for a surface and bulk, in contrast to the previous paper \cite {Borman2009} where the energy of vacancy formation was assumed to be mean quantity of a cluster height. Thus, according to the equation \eqref {eq:energ} the additional energy of a cluster of 30\AA in height due to the different formation energy of the vacancies for the surface and bulk is $ \left. {\delta E} \right | _ {h = 30\text {\AA}} \approx 0.1eV $.

\begin{figure}[H]
\centering
\includegraphics[width=0.7\linewidth]{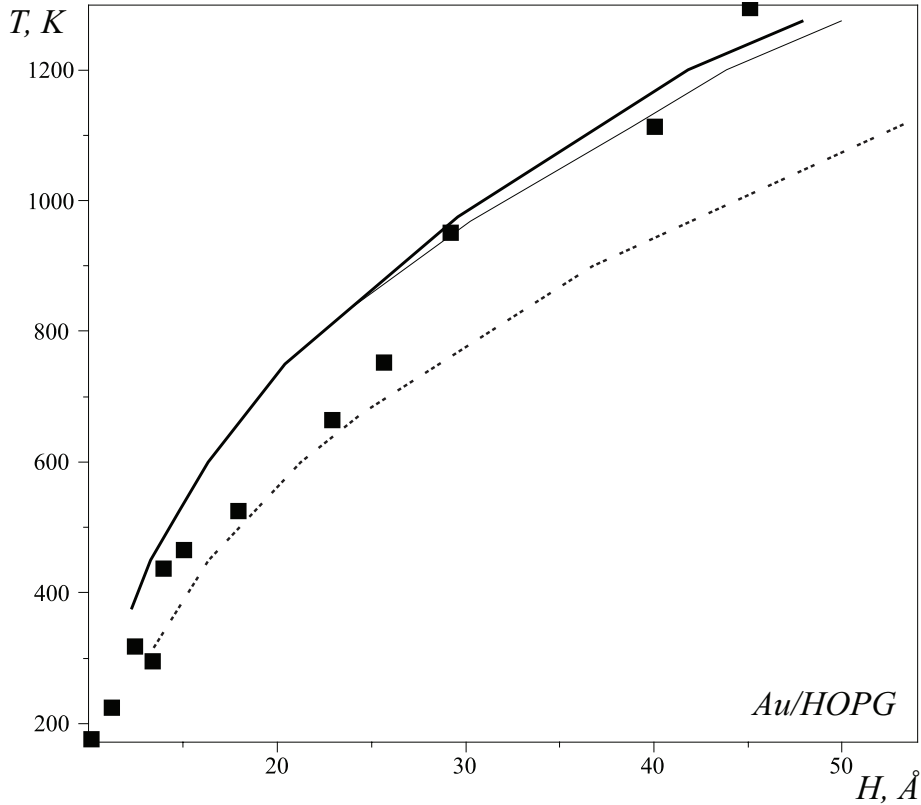}
\caption{\small{The dependences of the melting temperature  of Au cluster deposited onto HOPG surface. A solid curve --- with the account of the layer--by--layer melting, a dotted line is the dependence calculated without the account of the layer--by--layer melting, a dashed curve is the dependence calculated without the influence of a contact potential difference. Points \textemdash are the experimental data \cite{Borman2010}}} 
\label{fig:au_hopg_min}
\end{figure}

Fig.~\ref {fig:au_hopg_min} shows the dependences of melting temperature of Au clusters  deposited onto HOPG surface on the cluster height taking into account the layer-by-layer melting (a continuous line), without taking into account layer--by--layer melting (averaging energy of vacancy formation on cluster height) (a frequent dotted line) and without contact potential difference (a long dotted line). It is seen that taking into account the layer-by-layer melting results in significant changes in the curve behavior that quantitatively agrees with the experimental data. Taking into account  the contact potential difference of a cluster -- substrate changes a trend of curve insignificantly. Taking into account the finite lateral size of a cluster had not given appreciable changes in melting temperature of a cluster.

Thus, it is shown, the major effect on the shift of the melting point and the lattice parameter of supported nanoclusters under the high vacuum conditions is produced by the van der Waals cluster-substrate interaction. Taking into account the contact potential difference on a cluster--substrate interface, finite lateral cluster size, and  layer-by-layer melting of nanoparticles has been considered allowed one to satisfactorily describe the dependences of melting temperature on cluster height for Au and Ag clusters, deposited onto surfaces of W, HOPG, Ni.

 It is shown, that the account of layer--by--layer fusion of a cluster (i.e. observationally found phenomena when the melting of a cluster doesn't occur simultaneously in the whole volume of a cluster, but the presence both a liquid surface layer  and a solid core is possible) gives the considerable contribution to melting temperature of the clusters deposited onto various substrates that has allowed to one to satisfactorily describe the melting of nanoclusters of  various metals, deposited onto a different substrates.
 
 This work was supported by the Federal target program “Scientific and scientific-educational personnel of innovative Russia”.
\newpage 
\addcontentsline{toc}{chapter}{\tocsecindent{Литература}}
\bibliography{Bibliography_utf}

\end{document}